\documentclass[10pt]{article}
\usepackage{graphicx}
\usepackage{amsmath}
\usepackage{amssymb}
\usepackage{caption2}
\begin{document}
\begin{center}
{\large {\bf \sc{ The decay width of the  $Z_c(3900)$ as an  axialvector tetraquark state in solid quark-hadron duality
  }}} \\[2mm]
Zhi-Gang  Wang \footnote{E-mail: zgwang@aliyun.com.  }, Jun-Xia Zhang   \\
 Department of Physics, North China Electric Power University, Baoding 071003, P. R. China
\end{center}

\begin{abstract}
In this article, we tentatively assign the $Z_c^\pm(3900)$ to be  the diquark-antidiquark  type axialvector  tetraquark  state,   study the hadronic coupling  constants $G_{Z_cJ/\psi\pi}$, $G_{Z_c\eta_c\rho}$, $G_{Z_cD \bar{D}^{*}}$ with the QCD sum rules in details. We take into account both the connected and disconnected Feynman diagrams in carrying out the operator product expansion, as the connected Feynman diagrams alone cannot do the work.  Special attentions are paid to matching the hadron side of the correlation functions with the QCD side of the correlation functions to obtain solid duality, the routine can be applied to study other hadronic couplings directly. We study the two-body strong decays
$Z_c^+(3900)\to J/\psi\pi^+$, $\eta_c\rho^+$, $D^+ \bar{D}^{*0}$, $\bar{D}^0 D^{*+}$ and obtain the total width of the $Z_c^\pm(3900)$. The numerical results support assigning the $Z_c^\pm(3900)$ to be  the diquark-antidiquark  type axialvector  tetraquark  state, and assigning the $Z_c^\pm(3885)$ to be  the meson-meson  type axialvector  molecular  state.
\end{abstract}



\section{Introduction}

In 2013, the BESIII collaboration studied  the process  $e^+e^- \to \pi^+\pi^-J/\psi$ at a center-of-mass energy of 4.260 GeV using a $525\, \rm{pb}^{-1}$ data
sample collected with the BESIII detector, and observed a structure $Z_c(3900)$ in the $\pi^\pm J/\psi$ mass spectrum \cite{BES3900}. Then the structure $Z_c(3900)$ was confirmed by the Belle and CLEO collaborations \cite{Belle3900,CLEO3900}.
Also in 2013, the BESIII collaboration studied  the process $e^+e^- \to \pi D \bar{D}^*$, and observed a distinct charged structure $Z_c(3885)$  in the $(D \bar{D}^*)^{\pm}$
 mass spectrum  \cite{BES-3885}. The  angular distribution of the $\pi Z_c(3885)$ system favors a $J^P=1^+$ assignment \cite{BES-3885}.
 Furthermore, the BESIII collaboration measured the ratio $ R_{exp}$ \cite{BES-3885},
 \begin{eqnarray}
 R_{exp}&=&\frac{\Gamma(Z_c(3885)\to D\bar{D}^*)}{\Gamma(Z_c(3900)\to J/\psi \pi)} =6.2 \pm 1.1 \pm 2.7 \, .
 \end{eqnarray}
 In 2015, the BESIII collaboration observed the neutral parter $Z_c^0(3900)$  with a significance of $10.4\, \sigma$ in the process $e^+e^- \to \pi^0\pi^0J/\psi$ \cite{BESZ0-3900}. Recently, the BESIII collaboration determined   the  spin and parity of the $Z_c^\pm(3900)$ state to  be $J^P = 1^+$ with a statistical
significance larger than $7\sigma$ over other quantum numbers in a partial wave analysis of the process
$e^+e^- \to \pi^+\pi^-J/\psi$ \cite{JP-BES-Zc3900}.

Now we list out the  mass and width from different measurements.
\begin{flalign}
 & Z_c^\pm(3900) : M = 3899.0\pm 3.6\pm 4.9 \mbox{ MeV}\, , \, \Gamma = 46\pm 10\pm 20 \mbox{ MeV} \, , \,\,\, {\rm BESIII} \,\,\cite{BES3900} \, ,\nonumber \\
 & Z_c^\pm(3900) : M = 3894.5\pm 6.6\pm 4.5 \mbox{ MeV}\, , \, \Gamma = 63\pm 24 \pm 26 \mbox{ MeV} \, , \,\,\, {\rm Belle} \,\,\cite{Belle3900} \, , \nonumber \\
 & Z_c^\pm(3900) : M = 3886 \pm 4 \pm 2 \mbox{ MeV}\, , \, \Gamma = 37 \pm 4 \pm 8 \mbox{ MeV} \, , \,\,\, {\rm CLEO} \,\,\cite{CLEO3900} \, , \nonumber \\
 & Z_c^\pm(3885) : M = 3883.9 \pm 1.5 \pm 4.2 \mbox{ MeV}\, , \, \Gamma = 24.8 \pm 3.3 \pm 11.0 \mbox{ MeV} \, , \,\,\, {\rm BESIII} \,\,\cite{BES-3885} \, ,\nonumber \\
& Z_c^0(3900) : M = 3894.8 \pm 2.3 \pm 3.2 \mbox{ MeV}\, , \, \Gamma = 29.6 \pm 8.2 \pm 8.2 \mbox{ MeV} \, , \,\,\, {\rm BESIII} \,\,\cite{BESZ0-3900} \, .
\end{flalign}
The values of the mass are consistent with each other from  different measurements, while the values of the width differ from each other greatly. The $Z_c(3900)$ and $Z_c(3885)$ may be the same particle according to the mass, spin and parity.

 R. Faccini et al tentatively assign the $Z_c(3900)$ to be  the negative charge conjunction partner of the $X(3872)$ \cite{Maiani1303}. There have been several possible  assignments,
 such as tetraquark state \cite{Tetraquark3900,WangHuangTao-3900,Nielsen3900}, molecular state \cite{Molecular3900,WangMolecule-3900},  hadro-charmonium \cite{hadro-charmonium-3900}, rescattering effect \cite{FSI3900}.

In Ref.\cite{WangHuangTao-3900}, we study the masses and pole residues of the $J^{PC}=1^{+\pm}$ hidden charm tetraquark states with the QCD sum rules by calculating
 the contributions of the vacuum condensates up to dimension-10  in a consistent way in the operator product expansion,
and explore the energy scale dependence  in details for the first time. The predicted masses $M_{X}=3.87^{+0.09}_{-0.09}\,\rm{GeV}$ and
$M_{Z}=3.91^{+0.11}_{-0.09}\,\rm{GeV}$ support assigning    the $X(3872)$ and $Z_c(3900)$  to be the $1^{++}$
and $1^{+-}$ diquark-antidiquark type tetraquark states, respectively.

In Ref.\cite{WangMolecule-3900},  we study  the axialvector  hidden charm and hidden bottom molecular states with the QCD sum rules by calculating the  vacuum condensates up to dimension-10  in the operator product expansion, and explore the energy scale dependence of the QCD sum rules for the heavy molecular states  in details.  The numerical results  support assigning
  the $X(3872)$, $Z_c(3900)$, $Z_b(10610)$ to be the color singlet-singlet type  molecular states  with $J^{PC}=1^{++}$, $1^{+-}$, $1^{+-}$, respectively.

We can reproduce the experimental value of the mass of the $Z_c(3900)$ based on the QCD sum rules both in the scenario of tetraquark  states and in the scenario of molecule  states \cite{WangHuangTao-3900,WangMolecule-3900}. Additional theoretical works on the width are still needed to identify the $Z_c(3900)$.

In Ref.\cite{Nielsen3900},   Dias et al identify the $Z_c^\pm(3900)$   as the charged partner of the $X(3872)$ state, and study the two-body strong decays
$Z_c^+(3900)\to J/\psi\pi^+$, $\eta_c\rho^+$, $D^+ \bar{D}^{*0}$, $D^0 \bar{D}^{*+}$ with the QCD sum rules by  evaluating the three-point correlation  functions and take into account only the connected Feynman Diagrams, and they obtain the width $\Gamma_{Z_c}=63.0 \pm 18.1\,\rm{MeV}$.

In Ref.\cite{Azizi-Zc3900-decay},   Agaev et al study the two-body strong decays
$Z_c^+(3900)\to J/\psi\pi^+$, $\eta_c\rho^+$  with the light-cone QCD sum rules by taking into account both the connected and disconnected Feynman Diagrams, and obtain the width $\Gamma_{Z_c}=\Gamma(Z_c^+(3900)\to J/\psi\pi^+)+\Gamma(Z_c^+(3900)\to \eta_c\rho^+)=65.7 \pm 10.6\,\rm{MeV}$.

It is interesting to know that the connected  Feynman Diagrams alone or the connected plus disconnected Feynman Diagrams lead to the same result \cite{Nielsen3900,Azizi-Zc3900-decay}. As far as the $X(5568)$ is concerned, if we take the scenario of tetraquark  states, the width can also be reproduced  based on the connected  Feynman Diagrams alone \cite{Nielsen-5568} or the connected plus disconnected Feynman Diagrams \cite{Wang-5568,Azizi-5568}.  We should prove that the contributions of the disconnected Feynman diagrams can be neglected safely.

In this article, we assign the $Z_c(3900)$  to be the  diquark-antidiquark type tetraquark state with $J^{PC}=1^{+-}$,  study the hadronic coupling constants $G_{Z_cJ/\psi\pi}$, $G_{Z_c\eta_c\rho}$, $G_{Z_cD \bar{D}^{*}}$ with  the three-point QCD sum rules by   including both the connected and disconnected Feynman diagrams, special attentions  are paid to  the hadronic  spectral densities of the three-point correlation functions,   then calculate the partial decay widths of the strong decays  $Z_c^+(3900)\to J/\psi\pi^+$, $\eta_c\rho^+$, $D^+ \bar{D}^{*0}$, $D^0 \bar{D}^{*+}$, and diagnose the nature of the $Z_c^\pm(3900)$ based on the width and the ratio $R_{exp}=6.2 \pm 1.1 \pm 2.7$, if the $Z_c(3900)$ and $Z_c(3885)$ are the same particle with the diquark-antidiquark type structure.

The article is arranged as follows:  we derive the QCD sum rules for the hadronic coupling constants $G_{Z_cJ/\psi\pi}$, $G_{Z_c\eta_c\rho}$, $G_{Z_cD \bar{D}^{*}}$ in section 2; in Sect.3, we present the numerical results and discussions; and Sect.4 is reserved for our
conclusion.

\section{The width of the $Z_c(3900)$ as an axialvector tetraquark state }

We  study the two-body strong decays $Z_c^+(3900)\to J/\psi\pi^+$, $\eta_c\rho^+$, $D^+ \bar{D}^{*0}$, $\bar{D}^0 D^{*+}$ with the following three-point correlation functions $\Pi_{\mu\nu}^{1}(p,q)$, $\Pi_{\mu\nu}^{2}(p,q)$ and $\Pi_{\mu\nu}^{3}(p,q)$, respectively,
\begin{eqnarray}
\Pi_{\mu\nu}^{1}(p,q)&=&i^2\int d^4xd^4y \, e^{ipx}e^{iqy}\, \langle 0|T\left\{J_\mu^{J/\psi}(x)J_5^{\pi}(y)J_{\nu}(0)\right\}|0\rangle\, ,   \\
\Pi_{\mu\nu}^{2}(p,q)&=&i^2\int d^4xd^4y\, e^{ipx}e^{iqy}\, \langle 0|T\left\{J_5^{\eta_c}(x)J_\mu^{\rho}(y)J_{\nu}(0)\right\}|0\rangle \, , \\
\Pi_{\mu\nu}^{3}(p,q)&=&i^2\int d^4xd^4y \, e^{ipx}e^{iqy}\, \langle 0|T\left\{J_\mu^{D^*}(x)J_5^{D}(y)J_{\nu}(0)\right\}|0\rangle \, ,
\end{eqnarray}
where the currents
\begin{eqnarray}
J_\mu^{J/\psi}(x)&=&\bar{c}(x)\gamma_\mu c(x) \, ,\nonumber \\
J_5^{\pi}(y)&=&\bar{u}(y)i\gamma_5 d(y) \, ,  \\
J_5^{\eta_c}(x)&=&\bar{c}(x)i\gamma_5 c(x) \, ,\nonumber \\
J_\mu^{\rho}(y)&=&\bar{u}(y)\gamma_\mu d(y) \, ,  \\
J_\mu^{D^*}(x)&=&\bar{u}(x)\gamma_\mu c(x) \, ,\nonumber \\
J_5^{D}(y)&=&\bar{c}(y)i\gamma_5 d(y) \, , \\
J_\nu(0)&=&\frac{\varepsilon^{ijk}\varepsilon^{imn}}{\sqrt{2}}\left\{c^n(0)C\gamma_\nu u^m(0) \bar{c}^k(0)\gamma_5 C \bar{d}^j(0)-c^n(0)C\gamma_5 u^m(0)\bar{c}^k(0)\gamma_\nu C \bar{d}^j(0) \right\} \, ,
\end{eqnarray}
interpolate the mesons $J/\psi$, $\pi$, $\eta_c$, $\rho$, $D^*$, $D$ and $Z_c(3900)$, respectively.

We insert  a complete set of intermediate hadronic states with
the same quantum numbers as the current operators into the three-point
correlation functions $\Pi_{\mu\nu}^{1}(p,q)$, $\Pi_{\mu\nu}^{2}(p,q)$ and $\Pi_{\mu\nu}^{3}(p,q)$ \cite{SVZ79,Reinders85}, and  isolate the ground state
contributions to obtain the following results,
\begin{eqnarray}
\Pi_{\mu\nu}^{1}(p,q)&=& \frac{f_{\pi}M_{\pi}^2f_{J/\psi}M_{J/\psi}\lambda_{Z_c}G_{Z_cJ/\psi \pi}}{m_u+m_d} \frac{-i}{(M_{Z_c}^2-p^{\prime2})(M_{J/\psi}^2-p^2)(M_{\pi}^2-q^2)} \left(-g_{\mu\alpha}+\frac{p_{\mu}p_{\alpha}}{p^2} \right) \nonumber\\
&&\left(-g_{\nu}{}^{\alpha}+\frac{p^{\prime}_{\nu}p^{\prime\alpha}}{p^{\prime2}} \right)+\cdots  \nonumber\\
&=&\left\{ \frac{f_{\pi}M_{\pi}^2f_{J/\psi}M_{J/\psi}\lambda_{Z_c}G_{Z_cJ/\psi \pi}}{m_u+m_d} \frac{-i}{(M_{Z_c}^2-p^{\prime2})(M_{J/\psi}^2-p^2)(M_{\pi}^2-q^2)}\right.\nonumber\\
&&+ \frac{-i}{(M_{Z_c}^2-p^{\prime2})(M_{J/\psi}^2-p^2)} \int_{s^0_\pi}^\infty dt\frac{\rho_{Z_c\pi^\prime}(p^{\prime 2},p^2,t)}{t-q^2}\nonumber\\
&&+ \frac{-i}{(M_{Z_c}^2-p^{\prime2})(M_{\pi}^2-q^2)} \int_{s^0_{J/\psi}}^\infty dt\frac{\rho_{Z_c \psi^\prime}(p^{\prime2},t,q^2)}{t-p^2}\nonumber\\
&&\left.+ \frac{-i}{(M_{J/\psi}^2-p^{2})(M_{\pi}^2-q^2)} \int_{s^0_{Z_c}}^\infty dt\frac{\rho_{Z_c^\prime J/\psi}(t,p^2,q^2)+\rho_{Z_c^\prime \pi}(t,p^2,q^2)}{t-p^{\prime2}}+\cdots\right\}\left(g_{\mu\nu}+\cdots\right) +\cdots\nonumber\\
&=& \Pi_1(p^{\prime2},p^2,q^2)\,g_{\mu\nu}+\cdots\, ,
\end{eqnarray}

\begin{eqnarray}
\Pi_{\mu\nu}^{2}(p,q)&=& \frac{f_{\eta_c}M_{\eta_c}^2f_{\rho}M_{\rho}\lambda_{Z_c}G_{Z_c\eta_c \rho}}{2m_c} \frac{-i}{(M_{Z_c}^2-p^{\prime2})(M_{\eta_c}^2-p^2)(M_{\rho}^2-q^2)} \left(-g_{\mu\alpha}+\frac{q_{\mu}q_{\alpha}}{q^2} \right) \nonumber\\
&&\left(-g_{\nu}{}^{\alpha}+\frac{p^{\prime}_{\nu}p^{\prime\alpha}}{p^{\prime2}} \right)+\cdots  \nonumber\\
&=&\left\{ \frac{f_{\eta_c}M_{\eta_c}^2f_{\rho}M_{\rho}\lambda_{Z_c}G_{Z_c\eta_c \rho}}{2m_c} \frac{-i}{(M_{Z_c}^2-p^{\prime2})(M_{\eta_c}^2-p^2)(M_{\rho}^2-q^2)}\right.\nonumber\\
&&+  \frac{-i}{(M_{Z_c}^2-p^{\prime2})(M_{\eta_c}^2-p^2)}\int_{s^0_\rho}^\infty dt \frac{\rho_{Z_c\rho^\prime}(p^{\prime 2},p^2,t)}{t-q^2} \nonumber\\
&&+ \frac{-i}{(M_{Z_c}^2-p^{\prime2})(M_{\rho}^2-q^2)} \int_{s^0_{\eta_c}}^\infty dt\frac{\rho_{Z_c\eta_c^\prime}
(p^{\prime 2},t,q^2)}{t-p^2}\nonumber\\
&&\left.+ \frac{-i}{(M_{\eta_c}^2-p^{2})(M_{\rho}^2-q^2)} \int_{s^0_{Z_c}}^\infty dt\frac{\rho_{Z_c^\prime \eta_c}
(t,p^2,q^2)+\rho_{Z_c^\prime \rho}(t,p^2,q^2)}{t-p^{\prime2}}+\cdots\right\}\left(g_{\mu\nu}+\cdots\right) +\cdots\nonumber\\
&=& \Pi_2(p^{\prime2},p^2,q^2)\,g_{\mu\nu}+\cdots\, ,
\end{eqnarray}

\begin{eqnarray}
\Pi_{\mu\nu}^{3}(p,q)&=& \frac{f_{D}M_{D}^2f_{D^*}M_{D^*}\lambda_{Z_c}G_{Z_c D\bar{D}^*}}{m_c} \frac{-i}{(M_{Z_c}^2-p^{\prime2})(M_{D^*}^2-p^2)(M_{D}^2-q^2)} \left(-g_{\mu\alpha}+\frac{p_{\mu}p_{\alpha}}{p^2} \right) \nonumber\\
&&\left(-g_{\nu}{}^{\alpha}+\frac{p^{\prime}_{\nu}p^{\prime\alpha}}{p^{\prime2}} \right)+\cdots  \nonumber\\
&=&\left\{ \frac{f_{D}M_{D}^2f_{D^*}M_{D^*}\lambda_{Z_c}G_{Z_cD\bar{D}^*}}{m_c} \frac{-i}{(M_{Z_c}^2-p^{\prime2})(M_{D^*}^2-p^2)(M_{D}^2-q^2)} \  \right.\nonumber\\
&&+ \frac{-i}{(M_{Z_c}^2-p^{\prime2})(M_{D^*}^2-p^2)} \int_{s^0_D}^\infty dt\frac{\rho_{Z_cD^\prime}(p^{\prime 2},p^2,t)}{t-q^2}\nonumber\\
&&+ \frac{-i}{(M_{Z_c}^2-p^{\prime2})(M_{D}^2-q^2)} \int_{s^0_{D^*}}^\infty dt\frac{\rho_{Z_c D^{*\prime}}(p^{\prime2},t,q^2)}{t-p^2}\nonumber\\
&&\left.+ \frac{-i}{(M_{D^*}^2-p^{2})(M_{D}^2-q^2)} \int_{s^0_{Z_c}}^\infty dt\frac{\rho_{Z_c^\prime D^*}(t,p^2,q^2)+\rho_{Z_c^\prime D}(t,p^2,q^2)}{t-p^{\prime2}}+\cdots\right\}\left(g_{\mu\nu}+\cdots\right) +\cdots\nonumber\\
&=& \Pi_3(p^{\prime2},p^2,q^2)\,g_{\mu\nu}+\cdots \, ,
\end{eqnarray}
where $p^\prime=p+q$, the $f_{J/\psi}$, $f_{\pi}$, $f_{\eta_c}$, $f_{\rho}$, $f_{D^*}$, $f_{D}$ and $\lambda_{Z_c}$  are the decay constants of the mesons  $J/\psi$, $\pi$, $\eta_c$, $\rho$, $D^*$, $D$ and $Z_c(3900)$, respectively, the $G_{Z_cJ/\psi\pi}$, $G_{Z_c\eta_c\rho}$ and $G_{Z_c D\bar{D}^*}$ are the hadronic coupling constants, which are defined by
\begin{eqnarray}
\langle0|J_{\mu}^{J/\psi}(0)|J/\psi(p)\rangle&=&f_{J/\psi}M_{J/\psi}\,\xi_\mu \,\, , \nonumber \\
\langle0|J_{5}^{\pi}(0)|\pi(q)\rangle&=&\frac{f_{\pi}M_{\pi}^2}{m_u+m_d} \,\, ,  \\
\langle0|J_{5}^{\eta_c}(0)|\eta_c(p)\rangle&=&\frac{f_{\eta_c}M_{\eta_c}^2}{2m_c} \,\, , \nonumber \\
\langle0|J_{\mu}^{\rho}(0)|\rho(q)\rangle&=&f_{\rho}M_{\rho}\,\varepsilon_\mu \,\, , \nonumber \\
\langle0|J_{\mu}^{D^*}(0)|D^*(p)\rangle&=&f_{D^*}M_{D^*}\,\varsigma_\mu \,\, , \nonumber \\
\langle0|J_{5}^{D}(0)|D(q)\rangle&=&\frac{f_{D}M_{D}^2}{m_c} \,\, ,  \\
\langle Z_c(p^\prime)|J_\nu(0)|0\rangle&=&\lambda_{Z_c}\,\zeta_\nu^*\,\, \\
\langle J/\psi(p)\pi(q)|Z_c(p^{\prime})\rangle&=&\xi^*(p)\cdot\zeta(p^{\prime})\, G_{Z_cJ/\psi\pi} \, , \nonumber\\
\langle\eta_c(p)\rho(q)|Z_c(p^{\prime})\rangle&=&\varepsilon^*(q)\cdot\zeta(p^{\prime})\, G_{Z_c\eta_c\rho}  \, ,  \nonumber\\
\langle D^*(p)D(q)|Z_c(p^{\prime})\rangle&=&\varsigma^*(p)\cdot\zeta(p^{\prime})\, G_{Z_c D\bar{D}^*} \, ,
\end{eqnarray}
the $\xi$, $\varepsilon$, $\varsigma$ and $\zeta$ are polarization vectors of the $J/\psi$,  $\rho$, $D^*$ and $Z_c(3900)$, respectively. The
$s^0_{\pi}$, $s^0_{J/\psi}$, $s^0_{Z_c}$, $s^0_{\eta_c}$, $s^0_{\rho}$, $s^0_{D^*}$ and $s^0_{D}$ are the continuum threshold parameters.
The 12 unknown functions $ \rho_{Z_c\pi^\prime}(p^{\prime 2},p^2,t)$, $ \rho_{Z_c\psi^\prime}(p^{\prime 2},t,q^2)$,
$ \rho_{Z_c^\prime\pi}(t,p^2,q^2)$, $\rho_{Z_c^\prime J/\psi}(t,p^2,q^2)$, $\rho_{Z_c\rho^\prime}(p^{\prime 2},p^2,t)$,
$\rho_{Z_c\eta_c^\prime}(p^{\prime 2},t,q^2)$,
$\rho_{Z_c^\prime\rho}(t,p^2,q^2)$, $\rho_{Z_c^\prime \eta_c}(t,p^2,q^2)$, $\rho_{Z_c D^{\prime}}(p^{\prime 2},p^2,t)$,
$\rho_{Z_cD^{*\prime}}(p^{\prime 2},t,q^2)$, $\rho_{Z_c^\prime D^*}(t,p^2,q^2)$,
$\rho_{Z_c^\prime D}(t,p^2,q^2)$  have complex dependence on the transitions
between the ground states and the high resonances  or the continuum states.

In this article, we choose the tensor $g_{\mu\nu}$  to study the hadronic  coupling constants $G_{Z_cJ/\psi\pi}$, $G_{Z_c\eta_c\rho}$ and $G_{Z_c D^* D}$   to avoid the contaminations from the corresponding scalar and pseudoscalar mesons, as the following current-meson couplings are non-vanishing,
\begin{eqnarray}
\langle0|J_{\mu}^{J/\psi}(0)|\chi_{c0}(p)\rangle&=&f_{\chi_{c0}}p_\mu \,\, , \nonumber \\
\langle0|J_{\mu}^{\rho}(0)|a_0(q)\rangle&=&f_{a_0}q_\mu \,\, , \nonumber \\
\langle0|J_{\mu}^{D^*}(0)|D_0^*(p)\rangle&=&f_{D^*_0}p_\mu \,\, , \nonumber \\
\langle Z_{c0}(p^\prime)|J_\nu(0)|0\rangle&=&-i\,\lambda_{Z_{c0}}\,p^\prime_\nu\,\, ,
\end{eqnarray}
where the $f_{\chi_{c0}}$, $f_{a_0}$, $f_{D^*_0}$, $\lambda_{Z_{c0}}$ are the decay constants of the $\chi_{c0}(3414)$, $a_0(980)$, $D_0^*(2400)$ and $Z_{c}(J^P=0^-)$, respectively.  The terms  proportional to $p_\mu p^\prime_\nu$ in the $\Pi_{\mu\nu}^1(p,q)$ and $\Pi_{\mu\nu}^3(p,q)$ and the terms proportional to $q_\mu p^\prime_\nu$ in the $\Pi_{\mu\nu}^2(p,q)$ have contaminations from the hadronic coupling constants $G_{Z_c \chi_{c0}\pi}$, $G_{Z_c D^*_0 D}$ and $G_{Z_c\eta_c a_0}$, respectively.

 We introduce the notations $C_{Z_c\pi^\prime}$, $C_{Z_c\psi^\prime}$, $C_{Z_c^\prime\pi}$,
  $C_{Z_c^\prime J/\psi}$, $C_{Z_c\rho^\prime}$, $C_{Z_c\eta_c^\prime}$, $C_{Z_c^\prime\rho}$, $C_{Z_c^\prime \eta_c}$, $C_{Z_cD^{*\prime}}$, $C_{Z_cD^\prime}$,
   $C_{Z_c^\prime D^*}$ and $C_{Z_c^\prime D}$ to parameterize the net effects,
\begin{eqnarray}
C_{Z_c\pi^\prime}&=&\int_{s^0_\pi}^\infty dt\frac{ \rho_{Z_c\pi^\prime}(p^{\prime 2},p^2,t)}{t-q^2}\, ,\nonumber\\
C_{Z_c\psi^\prime}&=&\int_{s^0_{J/\psi}}^\infty dt\frac{\rho_{Z_c\psi^\prime}(p^{\prime 2},t,q^2)}{t-p^2}\, ,\nonumber\\
C_{Z_c^\prime\pi}&=&\int_{s^0_{Z_c}}^\infty dt\frac{ \rho_{Z_c^\prime\pi}(t,p^2,q^2)}{t-p^{\prime2}}\, ,\nonumber\\
C_{Z_c^\prime J/\psi}&=&\int_{s^0_{Z_c}}^\infty dt\frac{ \rho_{Z_c^\prime J/\psi}(t,p^2,q^2)}{t-p^{\prime2}}\, ,
\end{eqnarray}
\begin{eqnarray}
C_{Z_c\rho^\prime}&=&\int_{s^0_{\rho}}^\infty dt \frac{\rho_{Z_c\rho^\prime}(p^{\prime 2},p^2,t)}{t-q^2}\, ,\nonumber\\
C_{Z_c\eta_c^\prime}&=&\int_{s^0_{\eta_c}}^\infty dt \frac{ \rho_{Z_c\eta_c^\prime}(p^{\prime 2},t,q^2)}{t-p^2}\, , \nonumber\\
C_{Z_c^\prime\rho}&=&\int_{s^0_{Z_c}}^\infty dt\frac{ \rho_{Z_c^\prime\rho}(t,p^2,q^2)}{t-p^{\prime2}}\, ,\nonumber\\
C_{Z_c^\prime \eta_c}&=&\int_{s^0_{Z_c}}^\infty dt\frac{ \rho_{Z_c^\prime \eta_c}(t,p^2,q^2)}{t-p^{\prime2}}\, ,
\end{eqnarray}
\begin{eqnarray}
C_{Z_cD^{*\prime}}&=&\int_{s^0_{D^*}}^\infty dt \frac{\rho_{Z_c D^{*\prime}}(p^{\prime 2},t,q^2)}{t-p^2}\, ,\nonumber\\
C_{Z_cD^\prime}&=&\int_{s^0_{D}}^\infty dt \frac{ \rho_{Z_cD^\prime}(p^{\prime 2},p^2,t)}{t-q^2}\, , \nonumber\\
C_{Z_c^\prime D^*}&=&\int_{s^0_{Z_c}}^\infty dt\frac{ \rho_{Z_c^\prime D^*}(t,p^2,q^2)}{t-p^{\prime2}}\, ,\nonumber\\
C_{Z_c^\prime D}&=&\int_{s^0_{Z_c}}^\infty dt\frac{ \rho_{Z_c^\prime D}(t,p^2,q^2)}{t-p^{\prime2}}\, .
\end{eqnarray}

Then the correlation functions on the phenomenological side can be written as
\begin{eqnarray}
\Pi_1(p^{\prime2},p^2,q^2)&=& \frac{f_{\pi}M_{\pi}^2f_{J/\psi}M_{J/\psi}\lambda_{Z_c}G_{Z_cJ/\psi \pi}}{m_u+m_d} \frac{-i}{(M_{Z_c}^2-p^{\prime2})(M_{J/\psi}^2-p^2)(M_{\pi}^2-q^2)}\nonumber\\
&&+ \frac{-i C_{Z_c\pi^\prime}}{(M_{Z_c}^2-p^{\prime2})(M_{J/\psi}^2-p^2)}
 + \frac{-i C_{Z_c\psi^\prime}}{(M_{Z_c}^2-p^{\prime2})(M_{\pi}^2-q^2)}   \nonumber\\
 &&+  \frac{-i C_{Z_c^\prime J/\psi }-i C_{Z_c^\prime \pi }}{(M_{J/\psi}^2-p^{2})(M_{\pi}^2-q^2)} +\cdots \, ,
\end{eqnarray}

\begin{eqnarray}
\Pi_2(p^{\prime2},p^2,q^2)&=& \frac{f_{\eta_c}M_{\eta_c}^2f_{\rho}M_{\rho}\lambda_{Z_c}G_{Z_c\eta_c \rho}}{2m_c} \frac{-i}{(M_{Z_c}^2-p^{\prime2})(M_{\eta_c}^2-p^2)(M_{\rho}^2-q^2)}\nonumber\\
&&+  \frac{-i C_{Z_c\rho^\prime}}{(M_{Z_c}^2-p^{\prime2})(M_{\eta_c}^2-p^2)}+ \frac{-i C_{Z_c\eta_c^\prime}}{(M_{Z_c}^2-p^{\prime2})(M_{\rho}^2-q^2)}   \nonumber\\
&&+   \frac{-i C_{Z_c^\prime\eta_c}-i C_{Z_c^\prime\rho}}{(M_{\eta_c}^2-p^{2})(M_{\rho}^2-q^2)} +\cdots \, ,
\end{eqnarray}

\begin{eqnarray}
\Pi_3(p^{\prime2},p^2,q^2)&=&\frac{f_{D}M_{D}^2f_{D^*}M_{D^*}\lambda_{Z_c}G_{Z_cD\bar{D}^*}}{m_c} \frac{-i}{(M_{Z_c}^2-p^{\prime2})(M_{D^*}^2-p^2)(M_{D}^2-q^2)} \nonumber\\
&&+ \frac{-i C_{Z_cD^\prime}}{(M_{Z_c}^2-p^{\prime2})(M_{D^*}^2-p^2)}
 + \frac{-i C_{Z_c D^{*\prime}}}{(M_{Z_c}^2-p^{\prime2})(M_{D}^2-q^2)}   \nonumber\\
 &&+  \frac{-i C_{Z_c^\prime D^* }-i C_{Z_c^\prime D }}{(M_{D^*}^2-p^{2})(M_{D}^2-q^2)} +\cdots \, .
\end{eqnarray}

In numerical calculations,   we smear  the dependencies of the  $C_{Z_c\pi^\prime}$, $C_{Z_c\psi^\prime}$, $C_{Z_c^\prime\pi}$,
  $C_{Z_c^\prime J/\psi}$, $C_{Z_c\rho^\prime}$, $C_{Z_c\eta_c^\prime}$, $C_{Z_c^\prime\rho}$, $C_{Z_c^\prime \eta_c}$, $C_{Z_cD^{*\prime}}$, $C_{Z_cD^\prime}$,
   $C_{Z_c^\prime D^*}$ and $C_{Z_c^\prime D}$ on the momentums $p^{\prime2}$, $p^2$, $q^2$, and take them  as free parameters, and choose the suitable values  to
eliminate the contaminations from the high resonances and continuum states to obtain the stable QCD sum rules with the variations of
the Borel parameters.

We carry out the operator product expansion up to the vacuum condensates of dimension 5 and neglect the tiny contributions of the gluon condensate.
On the QCD side, the correlation functions $\Pi_{1}(p^{\prime2},p^2,q^2)$ and $\Pi_{2}(p^{\prime2},p^2,q^2)$ can be written as
\begin{eqnarray}
\Pi_1(p^{\prime2},p^2,q^2)&=& \frac{i}{32\sqrt{2}\pi^4} \int_{4m_c^2}^{\infty} ds \frac{1}{s-p^2} \int_{0}^{\infty} du \frac{1}{u-q^2}\, u \left(s+2m_c^2 \right)\sqrt{1-\frac{4m_c^2}{s}} \nonumber\\
&& +\frac{i m_c\langle\bar{q} q\rangle}{4\sqrt{2}\pi^2} \int_{4m_c^2}^{\infty} ds \frac{1}{s-p^2}   \frac{p^{\prime2}-s-q^2}{q^2}\,  \sqrt{1-\frac{4m_c^2}{s}} \nonumber\\
&& +\frac{i m_c\langle\bar{q}g_s\sigma Gq\rangle}{16\sqrt{2}\pi^2} \int_{4m_c^2}^{\infty} ds \frac{1}{s-p^2}   \frac{p^{\prime2}-s-q^2}{q^4}\,  \sqrt{1-\frac{4m_c^2}{s}} \nonumber\\
&& +\frac{i m_c\langle\bar{q}g_s\sigma Gq\rangle}{48\sqrt{2}\pi^2} \frac{\partial}{\partial m_A^2}\int_{(m_A+m_c)^2}^{\infty} ds \frac{1}{s-p^2}   \frac{p^{\prime2}-s-q^2}{q^2}\,  \frac{\sqrt{\lambda(s,m_A^2,m_c^2)}}{s}\mid_{m_A \to m_c} \nonumber\\
&& -\frac{i m_c\langle\bar{q}g_s\sigma Gq\rangle}{16\sqrt{2}\pi^2} \int_{4m_c^2}^{\infty} ds \frac{1}{s-p^2}   \frac{p^{\prime2}-s-q^2}{q^4}\,  \sqrt{1-\frac{4m_c^2}{s}}\, ,
\end{eqnarray}

\begin{eqnarray}
\Pi_2(p^{\prime2},p^2,q^2)&=& -\frac{i}{32\sqrt{2}\pi^4} \int_{4m_c^2}^{\infty} ds \frac{1}{s-p^2} \int_{0}^{\infty} du \frac{1}{u-q^2}\, u s \, \sqrt{1-\frac{4m_c^2}{s}} \nonumber\\
&& -\frac{i m_c\langle\bar{q} q\rangle}{4\sqrt{2}\pi^2} \int_{4m_c^2}^{\infty} ds \frac{1}{s-p^2}   \frac{p^{\prime2}-s-q^2}{q^2}\,  \sqrt{1-\frac{4m_c^2}{s}} \nonumber\\
&& -\frac{i m_c\langle\bar{q}g_s\sigma Gq\rangle}{16\sqrt{2}\pi^2} \int_{4m_c^2}^{\infty} ds \frac{1}{s-p^2}   \frac{p^{\prime2}-s-q^2}{q^4}\,  \sqrt{1-\frac{4m_c^2}{s}} \nonumber\\
&& +\frac{i m_c\langle\bar{q}g_s\sigma Gq\rangle}{48\sqrt{2}\pi^2} \frac{\partial}{\partial m_A^2}\int_{(m_A+m_c)^2}^{\infty} ds \frac{1}{s-p^2}   \frac{p^{\prime2}-s-q^2}{q^2}\,  \frac{\sqrt{\lambda(s,m_A^2,m_c^2)}}{s}\mid_{m_A \to m_c} \nonumber\\
&& -\frac{i m_c\langle\bar{q}g_s\sigma Gq\rangle}{48\sqrt{2}\pi^2} \int_{4m_c^2}^{\infty} ds \frac{1}{s-p^2}   \frac{p^{\prime2}-s-q^2}{q^4}\,  \sqrt{1-\frac{4m_c^2}{s}}\, ,
\end{eqnarray}
where the last two terms originate from the Feynman diagrams where a quark pair $\bar{q}q$ absorbs a gluon emitted from other quark line.
The term
\begin{eqnarray}
 +\frac{i m_c\langle\bar{q}g_s\sigma Gq\rangle}{48\sqrt{2}\pi^2} \frac{\partial}{\partial m_A^2}\int_{(m_A+m_c)^2}^{\infty} ds \frac{1}{s-p^2}   \frac{p^{\prime2}-s-q^2}{q^2}\,  \frac{\sqrt{\lambda(s,m_A^2,m_c^2)}}{s}\mid_{m_A \to m_c}\, ,
\end{eqnarray}
in above equations  comes  from the connected Feynman diagrams, if we set $p^{\prime2}=p^2$, then it reduces to
\begin{eqnarray}
&&-\frac{i m_c\langle\bar{q}g_s\sigma Gq\rangle}{48\sqrt{2}\pi^2} \frac{\partial}{\partial m_A^2}\int_{(m_A+m_c)^2}^{\infty} ds \frac{1}{s-p^2}   \,  \frac{\sqrt{\lambda(s,m_A^2,m_c^2)}}{s}\mid_{m_A \to m_c} \nonumber\\
&& -\frac{i m_c\langle\bar{q}g_s\sigma Gq\rangle}{48\sqrt{2}\pi^2} \frac{\partial}{\partial m_A^2}\int_{(m_A+m_c)^2}^{\infty} ds \frac{1}{q^2}   \,  \frac{\sqrt{\lambda(s,m_A^2,m_c^2)}}{s}\mid_{m_A \to m_c}\, .
\end{eqnarray}
It has no contribution after performing  the double Borel transformation with respect to the variables $P^2=-p^2$ and $Q^2=-q^2$. It is more reasonable to performing the Borel transformation than taking the limit $q^2 \to 0$, as  we carry out the operator product expansion at the large spacelike region $Q^2=-q^2 \to \infty$. So the connected Feynman diagrams have no contributions in the correlation functions $\Pi_{1/2}(p^{\prime2},p^2,q^2)$, which are  in contrary to Refs.\cite{Nielsen3900,Nielsen-5568}, where only the connected Feynman diagrams have contributions and the limit $Q^2 \to 0$ is taken.

For the correlation function $\Pi_3(p^{\prime2},p^2,q^2)$, only the connected Feynman diagrams have contributions, we can set $p^{\prime2}=4p^2$ according to the relation  $M_{Z_c(3900)}\approx 2M_{D^*}$, the complex expression of the  correlation function $\Pi_3(p^{\prime2},p^2,q^2)$ can be reduced to a more simple form,
\begin{eqnarray}
\Pi_3(4p^{2},p^2,q^2)&=&\frac{i m_c\langle\bar{q}g_s\sigma Gq\rangle}{96\sqrt{2}\pi^2}\int_{m_c^2}^{\infty} ds \frac{1}{s-p^2} \frac{1}{q^2-m_c^2}\left( \frac{9}{2}-\frac{10m_c^2}{s}+\frac{3m_c^4}{2s^2}\right)   \nonumber\\
&&+\frac{i m_c\langle\bar{q}g_s\sigma Gq\rangle}{96\sqrt{2}\pi^2} \frac{1}{p^2-m_c^2}\int_{m_c^2}^{\infty} du \frac{1}{u-q^2} \left( \frac{9}{2}-\frac{8m_c^2}{u}+\frac{15m_c^4}{2u^2}\right)   \, .
\end{eqnarray}

In the limit $M_{\pi}^2 \to 0$, $M^2_{\rho}\to 0$, $M_{D}^2 \to 0$ and $m_c^2\to 0$, we maybe  expect to choose $Q^2=-q^2$ off-shell, and match the terms proportional to $\frac{1}{Q^2}$ in the limit $Q^2 \to 0$ on the
hadron side with the ones on  the QCD side to obtain QCD sum rules for the momentum dependent hadronic coupling constants $G_{Z_cJ/\psi\pi}(Q^2)$, $G_{Z_c\eta_c\rho}(Q^2)$, $G_{Z_cD \bar{D}^{*}}(Q^2)$, then extract the values to the mass-shell $Q^2=-M_{\pi}^2$, $-M_{\rho}^2$ or $-M_{D}^2$ to obtain the physical values \cite{Nielsen3900}. However, the approximations $M^2_{\rho}\to 0$, $M_{D}^2 \to 0$ and $m_c^2\to 0$ are rather crude, and we carry out the operator product expansion at the large space-like region $Q^2=-q^2 \to \infty$.  We prefer taking the imaginary parts of the correlation functions $\Pi_{1/2/3}(p^{\prime2},p^2,q^2)$  with respect to $q^2+i\epsilon$ through dispersion relation and obtain the physical hadronic  spectral densities,  then take the Borel transform with respect to the $Q^2$ to obtain the QCD sum rules for the physical hadronic coupling constants.

We have to be cautious in matching the QCD side with the hadron  side of the correlation functions $\Pi_{1/2/3}(p^{\prime 2},p^2,q^2)$, as there appears  the variable $p^{\prime2}=(p+q)^2$.
We rewrite the correlation functions $\Pi^H_{1/2/3}(p^{\prime 2},p^2,q^2)$ on the hadron  side into the following form through dispersion relation,
\begin{eqnarray}
\Pi_1^{H}(p^{\prime 2},p^2,q^2)&=&\int_{(M_{J/\psi}+M_{\pi})^2}^{s_{Z_c}^0}ds^\prime \int_{4m_c^2}^{s^0_{J/\psi}}ds \int_0^{u^0_{\pi}}du  \frac{\rho_1^H(s^\prime,s,u)}{(s^\prime-p^{\prime2})(s-p^2)(u-q^2)}+\cdots\, , \\
\Pi^H_2(p^{\prime 2},p^2,q^2)&=&\int_{(M_{\eta_c}+M_{\rho})^2}^{s_{Z_c}^0}ds^\prime \int_{4m_c^2}^{s^0_{\eta_c}}ds \int_0^{u^0_{\rho}}du  \frac{\rho_2^H(s^\prime,s,u)}{(s^\prime-p^{\prime2})(s-p^2)(u-q^2)}+\cdots\, , \\
\Pi^H_3(p^{\prime 2},p^2,q^2)&=&\int_{(M_{D^*}+M_{D})^2}^{s^0_{Z_c}}ds^\prime \int_{m_c^2}^{s^0_{D^*}}ds \int_{m_c^2}^{u^0_D}du  \frac{\rho_3^H(s^\prime,s,u)}{(s^\prime-p^{\prime2})(s-p^2)(u-q^2)}+\cdots\, ,
\end{eqnarray}
where the $\rho_{1/2/3}^H(s^\prime,s,u)$   are the hadronic spectral densities,
\begin{eqnarray}
\rho_{1/2/3}^H(s^\prime,s,u)&=&{\lim_{\epsilon_3\to 0}}\,\,{\lim_{\epsilon_2\to 0}} \,\,{\lim_{\epsilon_1\to 0}}\,\,\frac{ {\rm Im}_{s^\prime}\, {\rm Im}_{s}\,{\rm Im}_{u}\,\Pi^H_{1/2/3}(s^\prime+i\epsilon_3,s+i\epsilon_2,u+i\epsilon_1) }{\pi^3} \, ,
\end{eqnarray}
we add the superscript $H$ to denote the hadron side.
However, on the QCD side, the QCD spectral densities $\rho_{QCD}^{1/2/3}(s^\prime,s,u)$ do not exist,
\begin{eqnarray}
\rho_{QCD}^{1/2/3}(s^\prime,s,u)&=&{\lim_{\epsilon_3\to 0}}\,\,{\lim_{\epsilon_2\to 0}} \,\,{\lim_{\epsilon_1\to 0}}\,\,\frac{ {\rm Im}_{s^\prime}\, {\rm Im}_{s}\,{\rm Im}_{u}\,\Pi^{QCD}_{1/2/3}(s^\prime+i\epsilon_3,s+i\epsilon_2,u+i\epsilon_1) }{\pi^3} \nonumber\\
&=&0\, ,
\end{eqnarray}
because
\begin{eqnarray}
{\lim_{\epsilon_3\to 0}}\,\,\frac{ {\rm Im}_{s^\prime}\,\Pi^{QCD}_{1/2/3}(s^\prime+i\epsilon_3,p^2,q^2) }{\pi} &=&0\, ,
\end{eqnarray}
we add the superscript $QCD$ to denote the QCD side.

On the QCD side, the correlation functions $\Pi^{QCD}_{1/2/3}(p^{\prime 2},p^2,q^2)$ can be written into the following form through dispersion relation,
\begin{eqnarray}
\Pi^{QCD}_1(p^{\prime 2},p^2,q^2)&=&  \int_{4m_c^2}^{s^0_{J/\psi}}ds \int_0^{u^0_{\pi}}du  \frac{\rho^{QCD}_1(p^{\prime2},s,u)}{(s-p^2)(u-q^2)}+\cdots\, , \\
\Pi^{QCD}_2(p^{\prime 2},p^2,q^2)&=&  \int_{4m_c^2}^{s^0_{\eta_c}}ds \int_0^{u^0_{\rho}}du  \frac{\rho_2^{QCD}(p^{\prime 2},s,u)}{(s-p^2)(u-q^2)}+\cdots\, , \\
\Pi^{QCD}_3(p^{\prime 2},p^2,q^2)&=&  \int_{m_c^2}^{s^0_{D^*}}ds \int_{m_c^2}^{u^0_D}du  \frac{\rho_3^{QCD}(p^{\prime 2},s,u)}{(s-p^2)(u-q^2)}+\cdots\, ,
\end{eqnarray}
where the $\rho_{1/2/3}^{QCD}(p^{\prime 2},s,u)$   are the QCD spectral densities,
\begin{eqnarray}
\rho_{1/2/3}^{QCD}(p^{\prime 2},s,u)&=& {\lim_{\epsilon_2\to 0}} \,\,{\lim_{\epsilon_1\to 0}}\,\,\frac{  {\rm Im}_{s}\,{\rm Im}_{u}\,\Pi_{1/2/3}^{QCD}(p^{\prime 2},s+i\epsilon_2,u+i\epsilon_1) }{\pi^2} \, ,
\end{eqnarray}

We math the hadron side of the correlation functions  with the QCD side of the correlation functions,
 \begin{eqnarray}
  \int_{4m_c^2}^{s^0_{J/\psi}}ds \int_0^{u^0_{\pi}}du  \frac{\rho^{QCD}_1(p^{\prime2},s,u)}{(s-p^2)(u-q^2)}&=&\int_{(M_{J/\psi}+M_{\pi})^2}^{\infty}ds^\prime \int_{4m_c^2}^{s^0_{J/\psi}}ds \int_0^{u^0_{\pi}}du  \frac{\rho_1^H(s^\prime,s,u)}{(s^\prime-p^{\prime2})(s-p^2)(u-q^2)} \nonumber\\
  &=& \frac{f_{\pi}M_{\pi}^2f_{J/\psi}M_{J/\psi}\lambda_{Z_c}G_{Z_cJ/\psi \pi}}{m_u+m_d} \frac{-i}{(M_{Z_c}^2-p^{\prime2})(M_{J/\psi}^2-p^2)(M_{\pi}^2-q^2)}\nonumber\\
 &&+  \frac{-i C_{Z_c^\prime J/\psi }-i C_{Z_c^\prime \pi }}{(M_{J/\psi}^2-p^{2})(M_{\pi}^2-q^2)}  \, ,
\end{eqnarray}

\begin{eqnarray}
  \int_{4m_c^2}^{s^0_{\eta_c}}ds \int_0^{u^0_{\rho}}du  \frac{\rho_2^{QCD}(p^{\prime 2},s,u)}{(s-p^2)(u-q^2)}&=&\int_{(M_{\eta_c}+M_{\rho})^2}^{\infty}ds^\prime \int_{4m_c^2}^{s^0_{\eta_c}}ds \int_0^{u^0_{\rho}}du  \frac{\rho_2^H(s^\prime,s,u)}{(s^\prime-p^{\prime2})(s-p^2)(u-q^2)}\nonumber \\
 &=& \frac{f_{\eta_c}M_{\eta_c}^2f_{\rho}M_{\rho}\lambda_{Z_c}G_{Z_c\eta_c \rho}}{2m_c} \frac{-i}{(M_{Z_c}^2-p^{\prime2})(M_{\eta_c}^2-p^2)(M_{\rho}^2-q^2)}\nonumber\\
&&+   \frac{-i C_{Z_c^\prime\eta_c}-i C_{Z_c^\prime\rho}}{(M_{\eta_c}^2-p^{2})(M_{\rho}^2-q^2)} \, ,
\end{eqnarray}

\begin{eqnarray}
  \int_{m_c^2}^{s^0_{D^*}}ds \int_{m_c^2}^{u^0_D}du  \frac{\rho_3^{QCD}(p^{\prime 2},s,u)}{(s-p^2)(u-q^2)}&=&
  \int_{(M_{D^*}+M_{D})^2}^{\infty}ds^\prime \int_{m_c^2}^{s^0_{D^*}}ds \int_{m_c^2}^{u^0_D}du  \frac{\rho_3^H(s^\prime,s,u)}{(s^\prime-p^{\prime2})(s-p^2)(u-q^2)} \nonumber \\
&=&\frac{f_{D}M_{D}^2f_{D^*}M_{D^*}\lambda_{Z_c}G_{Z_cD\bar{D}^*}}{m_c} \frac{-i}{(M_{Z_c}^2-p^{\prime2})(M_{D^*}^2-p^2)(M_{D}^2-q^2)} \nonumber\\
 &&+  \frac{-i C_{Z_c^\prime D^* }-i C_{Z_c^\prime D }}{(M_{D^*}^2-p^{2})(M_{D}^2-q^2)}  \, ,
\end{eqnarray}
where the integrals over $ds^\prime$ are carried out firstly to obtain the solid duality,
\begin{eqnarray}
\int_{\Delta_s^2}^{s^0} ds \int_{\Delta_u^2}^{u^0} du \frac{\rho_{QCD}(p^{\prime2},s,u)}{(s-p^2)(u-q^2)}&=&\int_{\Delta_s^2}^{s^0} ds \int_{\Delta_u^2}^{u^0} du \frac{1}{(s-p^2)(u-q^2)}\left[ \int_{\Delta^2}^{\infty} ds^\prime \frac{\rho_{H}(p^{\prime2},s,u)}{s^\prime-p^{\prime2}}\right]\, , \nonumber\\
\end{eqnarray}
 the $\Delta_s^2$ and $\Delta_u^2$ denote the thresholds $4m_c^2$, $m_c^2$, $0$, the $\Delta^2$ denotes the thresholds $(M_{J/\psi}+M_{\pi})^2$, $(M_{\eta_c}+M_{\rho})^2$ and $(M_{D^*}+M_{D})^2$. No approximation is needed,  the continuum threshold parameter $s^0_{Z_c}$ in the $s^\prime$ channel is also not needed. The present  routine can be applied to study other hadronic couplings directly.

Then we set $p^{\prime2}=p^2$ and $p^{\prime2}=4p^2$ in the correlation functions $\Pi_{1/2}(p^{\prime 2},p^2,q^2)$ and $\Pi_{3}(p^{\prime 2},p^2,q^2)$, respectively, and perform the double Borel transformations with respect to the variables $P^2=-p^2$ and $Q^2=-q^2$, respectively  to obtain the following QCD sum rules,
\begin{eqnarray}
&&\frac{f_{\pi}M_{\pi}^2f_{J/\psi}M_{J/\psi}\lambda_{Z_c}G_{Z_cJ/\psi \pi}}{m_u+m_d}\frac{1}{M_{Z_c}^2-M_{J/\psi}^2} \left[ \exp\left(-\frac{M_{J/\psi}^2}{T^2} \right)-\exp\left(-\frac{M_{Z_c}^2}{T^2} \right)\right]\exp\left(-\frac{M_{\pi}^2}{T_2^2} \right) \nonumber\\
&&+\left[ C_{Z_c^\prime J/\psi}+C_{Z_c^\prime\pi}\right] \exp\left(-\frac{M_{J/\psi}^2}{T^2} -\frac{M_{\pi}^2}{T_2^2} \right)=-\frac{1}{32\sqrt{2}\pi^4}\int_{4m_c^2}^{s^0_{J/\psi}} ds \int_{0}^{u^0_{\pi}} du  u\left(s+2m_c^2\right)\sqrt{1-\frac{4m_c^2}{s}}\nonumber\\
&&\exp\left(-\frac{s}{T^2} -\frac{u}{T_2^2} \right)\, ,
\end{eqnarray}

\begin{eqnarray}
&&\frac{f_{\eta_c}M_{\eta_c}^2f_{\rho}M_{\rho}\lambda_{Z_c}G_{Z_c\eta_c \rho}}{2m_c } \frac{1}{M_{Z_c}^2-M_{\eta_c}^2}\left[ \exp\left(-\frac{M_{\eta_c}^2}{T^2} \right)-\exp\left(-\frac{M_{Z_c}^2}{T^2} \right)\right]\exp\left(-\frac{M_{\rho}^2}{T_2^2} \right) \nonumber\\
&&+\left[ C_{Z_c^\prime \eta_c}+C_{Z_c^\prime \rho}\right] \exp\left(-\frac{M_{\eta_c}^2}{T^2} -\frac{M_{\rho}^2}{T_2^2} \right)=\frac{1}{32\sqrt{2}\pi^4}\int_{4m_c^2}^{s^0_{\eta_c}} ds \int_{0}^{u^0_{\rho}} du  us\sqrt{1-\frac{4m_c^2}{s}}\exp\left(-\frac{s}{T^2} -\frac{u}{T_2^2} \right)\nonumber\\
&&+\frac{m_c\langle\bar{q}g_s\sigma Gq\rangle}{12\sqrt{2}\pi^2} \int_{4m_c^2}^{s^0_{\eta_c}} ds \sqrt{1-\frac{4m_c^2}{s}}\exp\left(-\frac{s}{T^2}  \right)\, ,
\end{eqnarray}

\begin{eqnarray}
&&\frac{f_{D}M_{D}^2f_{D^*}M_{D^*}\lambda_{Z_c}G_{Z_c D\bar{D}^*}}{4m_c } \frac{1}{\widetilde{M}_{Z_c}^2-M_{D^*}^2}\left[ \exp\left(-\frac{M_{D^*}^2}{T^2} \right)-\exp\left(-\frac{\widetilde{M}_{Z_c}^2}{T^2} \right)\right]\exp\left(-\frac{M_{D}^2}{T_2^2} \right) \nonumber\\
&&+\left[ C_{Z_c^\prime D^*}+C_{Z_c^\prime D}\right] \exp\left(-\frac{M_{D^*}^2}{T^2} -\frac{M_{D}^2}{T_2^2} \right)=\frac{m_c\langle\bar{q}g_s\sigma Gq\rangle}{96\sqrt{2}\pi^2}\int_{m_c^2}^{s^0_{D^*}} ds  \left( \frac{9}{2}-\frac{10m_c^2}{s}+\frac{3m_c^4}{2s^2}\right)   \nonumber\\
&&\exp\left(-\frac{s}{T^2} -\frac{m_c^2}{T_2^2} \right)+\frac{m_c\langle\bar{q}g_s\sigma Gq\rangle}{96\sqrt{2}\pi^2}\int_{m_c^2}^{u^0_{D}} du  \left( \frac{9}{2}-\frac{8m_c^2}{u}+\frac{15m_c^4}{2u^2}\right)   \exp\left(-\frac{m_c^2}{T^2} -\frac{u}{T_2^2} \right)\, ,
\end{eqnarray}	
where the  $s^0_{J/\psi}$, $u^0_{\pi}$,  $s^0_{\eta_c}$, $u^0_{\rho}$, $s^0_{D^*}$ and $u^0_{D}$ are the continuum threshold parameters, the $T^2$ and $T^2_2$ are the Borel parameters.

In the three QCD sum rules, the terms depend on $T^2_2$ can be factorized out explicitly,
\begin{eqnarray}
&&\frac{f_{\pi}M_{\pi}^2f_{J/\psi}M_{J/\psi}\lambda_{Z_c}G_{Z_cJ/\psi \pi}}{m_u+m_d}\frac{1}{M_{Z_c}^2-M_{J/\psi}^2} \left[ \exp\left(-\frac{M_{J/\psi}^2}{T^2} \right)-\exp\left(-\frac{M_{Z_c}^2}{T^2} \right)\right]  \nonumber\\
&&+\left[ C_{Z_c^\prime J/\psi}+C_{Z_c^\prime\pi}\right] \exp\left(-\frac{M_{J/\psi}^2}{T^2}  \right)=-\frac{1}{32\sqrt{2}\pi^4}\int_{4m_c^2}^{s^0_{J/\psi}} ds \int_{0}^{u^0_{\pi}} du  u\left(s+2m_c^2\right)\sqrt{1-\frac{4m_c^2}{s}}\nonumber\\
&&\exp\left(-\frac{s}{T^2} -\frac{u-M_{\pi}^2}{T_2^2} \right)\, ,
\end{eqnarray}

\begin{eqnarray}
&&\frac{f_{\eta_c}M_{\eta_c}^2f_{\rho}M_{\rho}\lambda_{Z_c}G_{Z_c\eta_c \rho}}{2m_c } \frac{1}{M_{Z_c}^2-M_{\eta_c}^2}\left[ \exp\left(-\frac{M_{\eta_c}^2}{T^2} \right)-\exp\left(-\frac{M_{Z_c}^2}{T^2} \right)\right] \nonumber\\
&&+\left[ C_{Z_c^\prime \eta_c}+C_{Z_c^\prime \rho}\right] \exp\left(-\frac{M_{\eta_c}^2}{T^2} \right)=\frac{1}{32\sqrt{2}\pi^4}\int_{4m_c^2}^{s^0_{\eta_c}} ds \int_{0}^{u^0_{\rho}} du  us\sqrt{1-\frac{4m_c^2}{s}}\exp\left(-\frac{s}{T^2} -\frac{u-M_{\rho}^2}{T_2^2} \right)\nonumber\\
&&+\frac{m_c\langle\bar{q}g_s\sigma Gq\rangle}{12\sqrt{2}\pi^2} \int_{4m_c^2}^{s^0_{\eta_c}} ds \sqrt{1-\frac{4m_c^2}{s}}\exp\left(-\frac{s}{T^2}+\frac{M_{\rho}^2}{T_2^2}  \right)\, ,
\end{eqnarray}

\begin{eqnarray}
&&\frac{f_{D}M_{D}^2f_{D^*}M_{D^*}\lambda_{Z_c}G_{Z_c D\bar{D}^*}}{4m_c } \frac{1}{\widetilde{M}_{Z_c}^2-M_{D^*}^2}\left[ \exp\left(-\frac{M_{D^*}^2}{T^2} \right)-\exp\left(-\frac{\widetilde{M}_{Z_c}^2}{T^2} \right)\right] \nonumber\\
&&+\left[ C_{Z_c^\prime D^*}+C_{Z_c^\prime D}\right] \exp\left(-\frac{M_{D^*}^2}{T^2}  \right)=\frac{m_c\langle\bar{q}g_s\sigma Gq\rangle}{96\sqrt{2}\pi^2}\int_{m_c^2}^{s^0_{D^*}} ds  \left( \frac{9}{2}-\frac{10m_c^2}{s}+\frac{3m_c^4}{2s^2}\right)   \nonumber\\
&&\exp\left(-\frac{s}{T^2} -\frac{m_c^2-M_{D}^2}{T_2^2} \right)+\frac{m_c\langle\bar{q}g_s\sigma Gq\rangle}{96\sqrt{2}\pi^2}\int_{m_c^2}^{u^0_{D}} du  \left( \frac{9}{2}-\frac{8m_c^2}{u}+\frac{15m_c^4}{2u^2}\right)   \exp\left(-\frac{m_c^2}{T^2} -\frac{u-M_{D}^2}{T_2^2} \right)\, , \nonumber\\
\end{eqnarray}	
the dependence on the Borel parameter $T^2_2$ is  trivial, $\exp\left(-\frac{u-M_{\pi}^2}{T_2^2} \right)$, $\exp\left(-\frac{u-M_{\rho}^2}{T_2^2} \right)$, $\exp\left(-\frac{u-M_{D}^2}{T_2^2} \right)$,  $\exp\left(-\frac{m_c^2-M_{D}^2}{T_2^2} \right)$, which differ from the QCD sum rules for the three-meson hadronic coupling constants greatly \cite{Nielsen-PPNP}. It is difficult to obtain $T_2^2$ independent regions in the present three QCD sum rules, as no other terms to stabilize the QCD sum rules.
We can take the local limit $T^2_2\to\infty$, which is so called  local-duality limit (the local QCD sum rules are reproduced from the   original
QCD sum rules in infinite Borel parameter limit) \cite{Local-QCDSR}, then $\exp\left(-\frac{u}{T_2^2} \right)=\exp\left(-\frac{m_c^2}{T_2^2} \right)=\exp\left(-\frac{M_\pi^2}{T_2^2} \right)=\exp\left(-\frac{M_\rho^2}{T_2^2} \right)=\exp\left(-\frac{M_D^2}{T_2^2} \right)=1$, the three QCD sum rules are greatly simplified.

Now we write down the simplified QCD sum rules explicitly,
\begin{eqnarray}
&&\frac{f_{\pi}M_{\pi}^2f_{J/\psi}M_{J/\psi}\lambda_{Z_c}G_{Z_cJ/\psi \pi}}{m_u+m_d}\frac{1}{M_{Z_c}^2-M_{J/\psi}^2} \left[ \exp\left(-\frac{M_{J/\psi}^2}{T^2} \right)-\exp\left(-\frac{M_{Z_c}^2}{T^2} \right)\right]  \nonumber\\
&&+\left[ C_{Z_c^\prime J/\psi}+C_{Z_c^\prime\pi}\right] \exp\left(-\frac{M_{J/\psi}^2}{T^2}  \right)=-\frac{1}{32\sqrt{2}\pi^4}\int_{4m_c^2}^{s^0_{J/\psi}} ds \int_{0}^{u^0_{\pi}} du  u\left(s+2m_c^2\right)\sqrt{1-\frac{4m_c^2}{s}}\nonumber\\
&&\exp\left(-\frac{s}{T^2} \right)\, ,
\end{eqnarray}

\begin{eqnarray}
&&\frac{f_{\eta_c}M_{\eta_c}^2f_{\rho}M_{\rho}\lambda_{Z_c}G_{Z_c\eta_c \rho}}{2m_c } \frac{1}{M_{Z_c}^2-M_{\eta_c}^2}\left[ \exp\left(-\frac{M_{\eta_c}^2}{T^2} \right)-\exp\left(-\frac{M_{Z_c}^2}{T^2} \right)\right] \nonumber\\
&&+\left[ C_{Z_c^\prime \eta_c}+C_{Z_c^\prime \rho}\right] \exp\left(-\frac{M_{\eta_c}^2}{T^2} \right)=\frac{1}{32\sqrt{2}\pi^4}\int_{4m_c^2}^{s^0_{\eta_c}} ds \int_{0}^{u^0_{\rho}} du  us\sqrt{1-\frac{4m_c^2}{s}}\exp\left(-\frac{s}{T^2}  \right)\nonumber\\
&&+\frac{m_c\langle\bar{q}g_s\sigma Gq\rangle}{12\sqrt{2}\pi^2} \int_{4m_c^2}^{s^0_{\eta_c}} ds \sqrt{1-\frac{4m_c^2}{s}}\exp\left(-\frac{s}{T^2} \right)\, ,
\end{eqnarray}

\begin{eqnarray}
&&\frac{f_{D}M_{D}^2f_{D^*}M_{D^*}\lambda_{Z_c}G_{Z_c D\bar{D}^*}}{4m_c } \frac{1}{\widetilde{M}_{Z_c}^2-M_{D^*}^2}\left[ \exp\left(-\frac{M_{D^*}^2}{T^2} \right)-\exp\left(-\frac{\widetilde{M}_{Z_c}^2}{T^2} \right)\right] \nonumber\\
&&+\left[ C_{Z_c^\prime D^*}+C_{Z_c^\prime D}\right] \exp\left(-\frac{M_{D^*}^2}{T^2}  \right)=\frac{m_c\langle\bar{q}g_s\sigma Gq\rangle}{96\sqrt{2}\pi^2}\int_{m_c^2}^{s^0_{D^*}} ds  \left( \frac{9}{2}-\frac{10m_c^2}{s}+\frac{3m_c^4}{2s^2}\right)\exp\left(-\frac{s}{T^2}  \right)   \nonumber\\
&&+\frac{m_c\langle\bar{q}g_s\sigma Gq\rangle}{96\sqrt{2}\pi^2}\int_{m_c^2}^{u^0_{D}} du  \left( \frac{9}{2}-\frac{8m_c^2}{u}+\frac{15m_c^4}{2u^2}\right)   \exp\left(-\frac{m_c^2}{T^2} \right)\, ,
\end{eqnarray}	
where $\widetilde{M}_{Z_c}^2=\frac{M_{Z_c}^2}{4}$.

\section{Numerical results and discussions}	
The input parameters on the QCD side are taken to be the standard values
$\langle\bar{q}q \rangle=-(0.24\pm 0.01\, \rm{GeV})^3$,
 $\langle\bar{q}g_s\sigma G q \rangle=m_0^2\langle \bar{q}q \rangle$,
$m_0^2=(0.8 \pm 0.1)\,\rm{GeV}^2$    at the energy scale  $\mu=1\, \rm{GeV}$
\cite{SVZ79,Reinders85,Colangelo-Review}, $m_{c}(m_c)=(1.28\pm0.03)\,\rm{GeV}$
 from the Particle Data Group \cite{PDG}. Furthermore, we set $m_u=m_d=0$ due to the small current quark masses.
 We take into account
the energy-scale dependence of  the input parameters from the renormalization group equation,
\begin{eqnarray}
\langle\bar{q}g_s \sigma Gq \rangle(\mu)&=&\langle\bar{q}g_s \sigma Gq \rangle(Q)\left[\frac{\alpha_{s}(Q)}{\alpha_{s}(\mu)}\right]^{\frac{2}{25}}\, , \nonumber\\
m_c(\mu)&=&m_c(m_c)\left[\frac{\alpha_{s}(\mu)}{\alpha_{s}(m_c)}\right]^{\frac{12}{25}} \, ,\nonumber\\
\alpha_s(\mu)&=&\frac{1}{b_0t}\left[1-\frac{b_1}{b_0^2}\frac{\log t}{t} +\frac{b_1^2(\log^2{t}-\log{t}-1)+b_0b_2}{b_0^4t^2}\right]\, ,
\end{eqnarray}
  where $t=\log \frac{\mu^2}{\Lambda^2}$, $b_0=\frac{33-2n_f}{12\pi}$, $b_1=\frac{153-19n_f}{24\pi^2}$, $b_2=\frac{2857-\frac{5033}{9}n_f+\frac{325}{27}n_f^2}{128\pi^3}$,  $\Lambda=210\,\rm{MeV}$, $292\,\rm{MeV}$  and  $332\,\rm{MeV}$ for the flavors  $n_f=5$, $4$ and $3$, respectively  \cite{PDG}, and evolve all the input parameters to the optimal energy scale  $\mu=1.4\,\rm{GeV}$ to extract hadronic coupling constants \cite{WangHuangTao-3900,Wang-1601}.

The hadronic parameters are taken as $M_{\pi}=0.13957\,\rm{GeV}$,  $M_{\rho}=0.77526\,\rm{GeV}$,
$M_{J/\psi}=3.0969\,\rm{GeV}$, $M_{\eta_c}=2.9834\,\rm{GeV}$ \cite{PDG},  $f_{\pi}=0.130\,\rm{GeV}$, $f_{\rho}=0.215\,\rm{GeV}$, $\sqrt{s^0_{\pi}}=0.85\,\rm{GeV}$, $\sqrt{s^0_{\rho}}=1.3\,\rm{GeV}$ \cite{Colangelo-Review},
$M_{D}=1.87\,\rm{GeV}$, $f_{D}=208\,\rm{MeV}$, $u^0_{D}=6.2\,\rm{GeV}^2$, $M_{D^*}=2.01\,\rm{GeV}$, $f_{D^*}=263\,\rm{MeV}$, $s^0_{D^*}=6.4\,\rm{GeV}^2$  \cite{WangJHEP},
$f_{J/\psi}=0.418 \,\rm{GeV}$, $f_{\eta_c}=0.387 \,\rm{GeV}$  \cite{Becirevic},  $\sqrt{s^0_{J/\psi}}=3.6\,\rm{GeV}$, $\sqrt{s^0_{\eta_c}}=3.5\,\rm{GeV}$, $M_{Z_c}=3.899\,\rm{GeV}$,   $\lambda_{Z_c}=2.1\times 10^{-2}\,\rm{GeV}^5$ \cite{WangHuangTao-3900,Wang-1601},  $f_{\pi}M^2_{\pi}/(m_u+m_d)=-2\langle \bar{q}q\rangle/f_{\pi}$ from the Gell-Mann-Oakes-Renner relation.

In the  scenario of tetraquark  states, the QCD sum rules indicate that the $Z_c(3900)$ and $Z(4430)$ can be tentatively assigned to be the ground state and the first radial excited state of the axialvector tetraquark states, respectively \cite{Wang4430}, the coupling of the current $J_\nu(0)$ to the excited state $Z(4430)$ is rather large, so the unknown parameters cannot be neglected.
The unknown parameters are fitted to be  $ C_{Z_c^\prime J/\psi}+C_{Z_c^\prime\pi}=0.001\,\rm{GeV}^8$, $C_{Z_c^\prime \eta_c}+C_{Z_c^\prime\rho}=0.0046\,\rm{GeV}^8 $
and $C_{Z_c^\prime D^*}+C_{Z_c^\prime D}=0.00013\,\rm{GeV}^8 $ to obtain  platforms in the Borel windows $T^2=(1.9-2.6)\,\rm{GeV}^2$, $(1.9-2.5)\,\rm{GeV}^2$  and $(1.5-2.1)\,\rm{GeV}^2$ for the hadronic coupling  constants $G_{Z_cJ/\psi\pi}$, $G_{Z_c\eta_c\rho}$, $G_{Z_cD \bar{D}^{*}}$, respectively.

\begin{figure}
 \centering
 \includegraphics[totalheight=5cm,width=7cm]{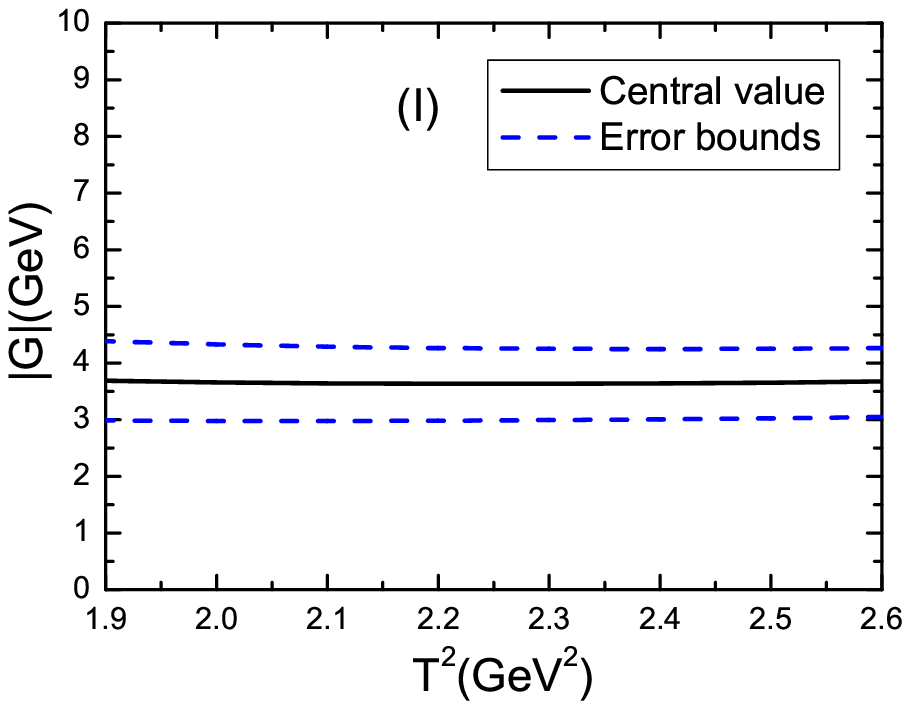}
 \includegraphics[totalheight=5cm,width=7cm]{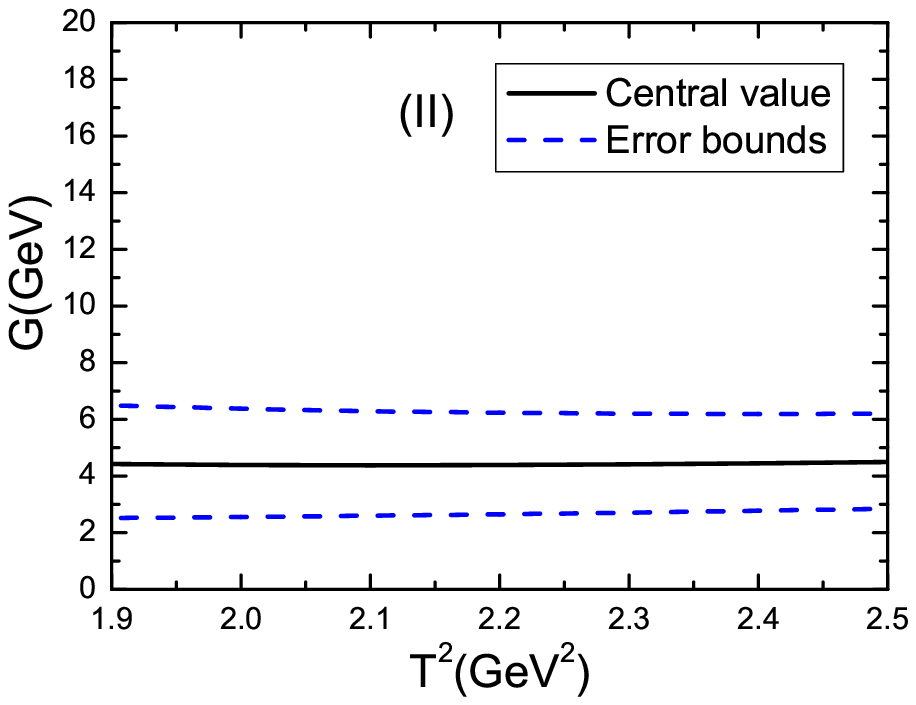}
 \includegraphics[totalheight=5cm,width=7cm]{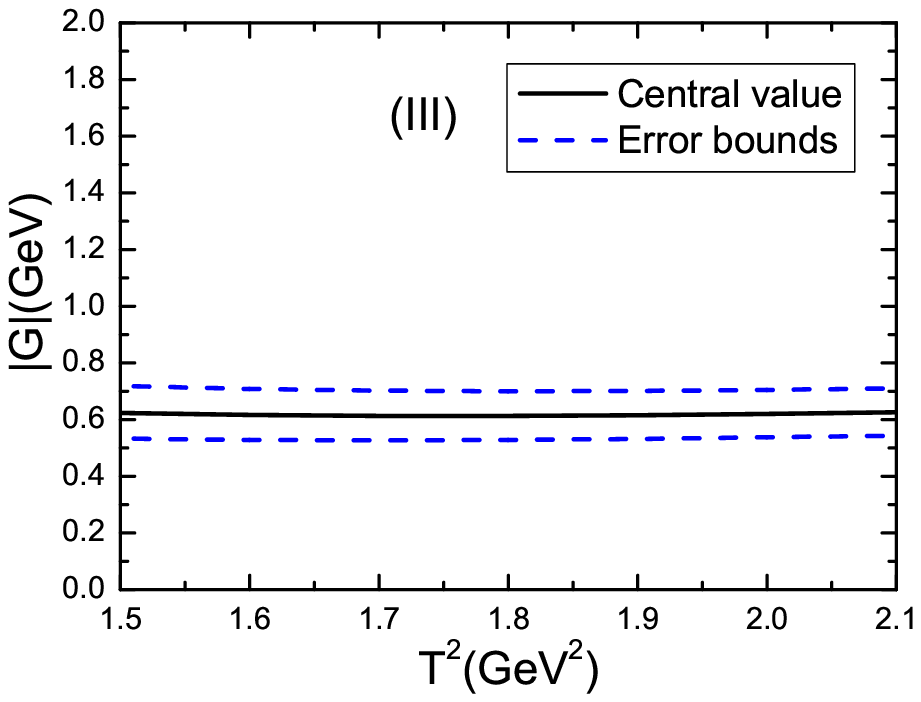}
        \caption{ The  hadronic coupling  constants $G_{Z_cJ/\psi\pi}$ (I), $G_{Z_c\eta_c\rho}$ (II) and $G_{Z_cD \bar{D}^{*}}$ (III) with variations of the Borel parameter $T^2$.  }
\end{figure}

Then it is easy to obtain the values of the hadronic coupling constants,
\begin{eqnarray}
|G_{Z_cJ/\psi \pi}| &=&3.63\pm0.70\,\rm{GeV}\, , \nonumber\\
G_{Z_c\eta_c\rho}&=&4.38\pm 1.86\,\rm{GeV}\, , \nonumber\\
|G_{Z_cD\bar{D}^*}|&=&0.62\pm0.09\,\rm{GeV}\, ,
\end{eqnarray}
which are shown explicitly in Fig.1.

We choose the masses  $M_{\pi}=0.13957\,\rm{GeV}$,  $M_{\rho}=0.77526\,\rm{GeV}$,
$M_{J/\psi}=3.0969\,\rm{GeV}$, $M_{\eta_c}=2.9834\,\rm{GeV}$,
$M_{D^+}=1.8695 \,\rm{GeV}$, $M_{D^{*0}}=2.00685\,\rm{GeV}$,  $M_{D^0}=1.86484 \,\rm{GeV}$, $M_{D^{*+}}=2.01026\,\rm{GeV}$ \cite{PDG},
  $M_{Z_c}=3.899\,\rm{GeV}$ \cite{BES3900}, and obtain the partial decay widths,
\begin{eqnarray}
\Gamma(Z_c^+(3900)\to J/\psi\pi^+)&=&25.8\pm 9.6\,\rm{MeV}   \, ,\nonumber\\
\Gamma(Z_c^+(3900)\to\eta_c\rho^+)&=&27.9\pm 20.1 \,\rm{MeV}  \, , \nonumber\\
\Gamma(Z_c^+(3900)\to D^+ \bar{D}^{*0})&=&0.22\pm 0.07\,\rm{MeV}   \, ,\nonumber\\
\Gamma(Z_c^+(3900)\to \bar{D}^0 D^{*+})&=&0.23\pm 0.07\,\rm{MeV}   \, ,
\end{eqnarray}
and the total width,
\begin{eqnarray}
\Gamma_{Z_c}&=&54.2\pm29.8\,\rm{MeV}\, ,
\end{eqnarray}
which is consistent with the experimental data considering the uncertainties \cite{BES3900,Belle3900,CLEO3900,BESZ0-3900}.
If we take the central values of the hadronic coupling constants $|G_{Z_cJ/\psi \pi}| =3.63\,\rm{GeV}$, $G_{Z_c\eta_c\rho}=4.38\,\rm{GeV}$,
 $|G_{Z_cD\bar{D}^*}|=0.62\,\rm{GeV}$, we can obtain the total width $\Gamma_{Z_c(3900)}=48.9\,\rm{MeV}$, which happens to coincide with the central value of the experimental dada $\Gamma = 46\pm 10\pm 20 \mbox{ MeV}$ from the BESIII collaboration \cite{BES3900}, while the predicted ratio
  \begin{eqnarray}
 R&=&\frac{\Gamma(Z_c(3900)\to D\bar{D}^*)}{\Gamma(Z_c(3900)\to J/\psi \pi)}=0.02\ll R_{exp} =\frac{\Gamma(Z_c(3885)\to D\bar{D}^*)}{\Gamma(Z_c(3900)\to J/\psi \pi)}=6.2 \pm 1.1 \pm 2.7 \, ,
 \end{eqnarray}
from  the BESIII collaboration \cite{BES-3885}. It is difficult to assign the $Z_c(3900)$ and $Z_c(3885)$ to be the same diquark-antidiquark type axialvector tetraquark state. We can assign the $Z_c(3900)$ to be the  diquark-antidiquark type axialvector tetraquark state, and assign the $Z_c^+(3885)$
to be  the molecular state $D^+\bar{D}^{*0}+D^{*+}\bar{D}^0$ according to the predicted mass $3.89\pm 0.09\,\rm{GeV}$ from the QCD sum rules \cite{WangMolecule-3900}. If the $Z_c(3885)$ is the $D^+\bar{D}^{*0}+D^{*+}\bar{D}^0$ molecular state, the decays to $D^+\bar{D}^{*0}$ and $D^{*+}\bar{D}^0$ take place through its component directly, it is easy to account for the large ratio $R_{exp}$.

Now we compare the present work with the work in Ref.\cite{Nielsen3900} in details. In the two works, the same currents are chosen except for the currents to interpolate the $\pi$ meson,  the operator product expansion is carried out at the large space-like regions $P^2=-p^2\rightarrow \infty$ and $Q^2=-q^2\rightarrow \infty$. In the present work, we take into account both the connected and disconnected Feynman diagrams, and obtain the solid quark-hadron duality by getting  the physical spectral densities through dispersion relation, then perform double Borel transforms  with respect to the variables $P^2$ and $Q^2$ to obtain the QCD sum rules for the physical hadronic coupling constants directly. We pay special attention to the hadron spectral spectral densities, and present detailed discussions and subtract
the continuum contaminations in a solid foundation. In Ref.\cite{Nielsen3900},  Dias et al take into account only the connected  Feynman diagrams, and obtain the quark-hadron duality by taking the limit $Q^2 \to 0$,  $M_{\pi}^2 \to 0$, $M^2_{\rho}\to 0$, $M_{D}^2 \to 0$ and $m_c^2\to 0$ and  choosing special tensor structures, then perform single Borel transform with respect to the variable $P^2$ to obtain the QCD sum rules for the momentum dependent hadronic coupling constants. They subtract the continuum contaminations by hand, then parameterize the momentum dependent hadronic coupling constants by some exponential functions with arbitrariness   to  extract the values to the mass-shell $Q^2=-M_{\pi}^2$, $-M_{\rho}^2$ or $-M_{D}^2$  to obtain the physical hadronic coupling constants. Although the values of the width of the $Z_c(3900)$ obtained in the present work and in Ref.\cite{Nielsen3900} are both compatible with the experimental data, the present predictions  have much less theoretical uncertainties.

\section{Conclusion}
In this article, we tentatively assign the $Z_c^\pm(3900)$ to be  the diquark-antidiquark  type axialvector  tetraquark  state,   study the hadronic coupling  constants $G_{Z_cJ/\psi\pi}$, $G_{Z_c\eta_c\rho}$, $G_{Z_cD \bar{D}^{*}}$ with the QCD sum rules in details. We introduce the three-point correlation functions, and carry out the operator product expansion up to the vacuum condensates of dimension-5, and neglect the tiny contributions of the gluon condensate.     In calculations, we take into account both the connected and disconnected Feynman diagrams, as the connected Feynman diagrams alone cannot do the work.   Special attentions are paid to matching the hadron side of the correlation functions with the QCD side of the correlation functions to obtain solid duality, the routine can be applied to study other hadronic couplings directly. We study the two-body strong decays
$Z_c^+(3900)\to J/\psi\pi^+$, $\eta_c\rho^+$, $D^+ \bar{D}^{*0}$, $\bar{D}^0 D^{*+}$ and obtain the total width of the $Z_c^\pm(3900)$, which is consistent with the experimental data. The numerical results support assigning the $Z_c^\pm(3900)$ to be  the diquark-antidiquark  type axialvector  tetraquark  state, and assigning the $Z_c^\pm(3885)$ to be  the meson-meson  type axialvector  molecular  state.

\section*{Acknowledgements}
This  work is supported by National Natural Science Foundation, Grant Number  11775079.

\end{document}